\documentclass[%
aip,
rsi,
reprint,
superscriptaddress,
nofootinbib,
 amsmath,amssymb,
]{revtex4-1}

\pdfoutput=1

\usepackage{graphicx}
\usepackage{amsmath,latexsym,tabularx}
\usepackage{xcolor}
\usepackage{multirow}
\usepackage[
  colorlinks=true,
  urlcolor=blue,
  linkcolor=blue,
  citecolor=blue
]{hyperref}
\usepackage{multirow,booktabs}
\usepackage{float}
\usepackage{upgreek}
\usepackage{textcomp}
\usepackage{braket}
\usepackage{siunitx}

\newcommand{\affilLL}[0]{Lincoln Laboratory, Massachusetts Institute of Technology, Lexington, Massachusetts 02421, USA}
\newcommand{\affilMIT}{Massachusetts Institute of Technology, Cambridge, Massachusetts 02139, USA}

\newcommand{\specificthanks}[1]{\@fnsymbol{#1}}
\begin{document}
\title{Cancellation of cryocooler-induced vibration in trapped-ion systems}

\author{Alexandra Plocki}
\altaffiliation{allie839@mit.edu}
\affiliation{\affilMIT}
\affiliation{\affilLL}

\author{Kyle DeBry}
\affiliation{\affilMIT}
\affiliation{\affilLL} 

\author{Agustin Valdes Martinez}
\altaffiliation{Present address: University of Chicago,
Chicago, Illinois, USA}
\affiliation{\affilMIT}
\affiliation{\affilLL}

\author{Isaac L. Chuang}
\affiliation{\affilMIT}

\author{Robert McConnell}
\affiliation{\affilLL}

\author{Colin D. Bruzewicz}
\affiliation{\affilLL}

\date{\today}

\begin{abstract}
  Mechanical cryocoolers offer numerous benefits for the operation of vacuum systems containing trapped atomic ions. However, the fidelity of laser-based gates can be degraded by phase noise induced by the cryocooler vibrations. We present a feed-forward vibration-cancellation system that directly measures inner shield displacement with an auxiliary interferometer and applies a compensating phase correction to the laser drive. The interferometer signal is demodulated, and a field-programmable gate array (FPGA) system unwraps the displacement measurement signal, computes the required correction phase, and updates the RF drive of an in-line acousto-optic modulator (AOM) to correct for the vibration-induced dephasing. With cancellation active, the qubit coherence time improves from $3.0\pm0.4$~ms to $35\pm5$~ms, consistent with a simulation based on the measured vibration spectrum. We further validate the cancellation by artificially increasing the mechanical coupling of the cryocooler to the vacuum chamber. The vibration-induced decoherence rate is suppressed by a factor of approximately 70 at an RMS (root mean square) displacement of 0.65~\textmu m.
\end{abstract}

\maketitle

\section{Introduction}
Trapped ions are a promising technology for scalable quantum information processing \cite{colin} and precision metrology,~\cite{brewer2019} owing to their long coherence times and well-isolated internal states. A common implementation uses a Gifford--McMahon (GM) cryocooler to cool a surface-electrode trap that confines the ions, offering long trap lifetimes and low motional heating rates by maintaining a trap temperature of a few kelvin.\cite{niedermayr2014} However, the operation of the GM cryocooler imparts vibration onto
the ions.~\cite{brandl2016, dubielzig2021, Spivey2022} For many laser-based gate operations, this vibration results in a time-varying frequency shift via the Doppler effect. The corresponding phase noise on the qubits degrades coherence and limits achievable laser gate fidelities.~\cite{schneider1998}

Previous experiments have characterized vibrations caused by cryocoolers using interferometry and reported displacement amplitudes ranging from \qty{7}{\nano\meter} to \qty{100}{\nano\meter}.\cite{Spivey2022, Qin2024, dubielzig2021, hartsell2026, brandl2016} These vibrations have been addressed through various stabilization
techniques, including passive vibration isolation,~\cite{Pagano2019, Micke2019} which reduces but does not eliminate coupling. Delivering the control light through photonics integrated into the trap chip makes the beams move in common with the ion, rendering the qubit coherence insensitive to cryocooler vibration,~\cite{niffenegger2020} but requires specialized trap fabrication. Alternatively, a co-propagating Raman beam geometry eliminates the ions'
sensitivity to vibration-induced phase errors, albeit at the cost of additional requirements on the beam path layout.\cite{Wineland1998, Lee2005}

Inspired by approaches that have been used to correct optical-fiber-induced phase noise,~\cite{Ma1994, Falke2012} we introduce a method to stabilize the laser phase based on interferometric measurements of the inner shield displacement. Here, in contrast to typical feedback-based fiber-noise cancellation,~\cite{Ma1994, Falke2012} we take a direct feed-forward approach.~\cite{Denecker2025, Hati2008, gouet2008, merlet2009, lautier2014} Specifically, we measure the inner shield displacement with a dedicated heterodyne interferometer and use that result to apply a phase shift to a gate-laser beam before it interacts with the ion.

This approach can be added to an existing apparatus with a minor change to an inner shield window and commercially available optical components. We achieve an order-of-magnitude improvement in the qubit coherence time, which is then limited by nonvibrational sources, and approximately 70-fold suppression of the vibration-induced decoherence rate at an RMS displacement of 0.65~\textmu m. The remainder of this manuscript is arranged as follows. Section~\ref{apparatus} describes the interferometer design, the FPGA-based signal-processing chain, and the measurements of the inner shield displacement. Section~\ref{experiment} describes the qubit system, the state-preparation protocol that maintains an isolated two-level system in $^{40}$Ca$^+$, and the Ramsey coherence-time measurement used to quantify performance. Section \ref{results} presents the measured coherence time with and without cancellation and the suppression of the vibration-induced decoherence rate as a function of applied vibration amplitude. We conclude in Section~\ref{discussion} with a discussion of remaining limitations and prospects for further improvement.

\section{Apparatus}\label{apparatus}
\begin{figure*}
    \centering
    \includegraphics[width=\linewidth]{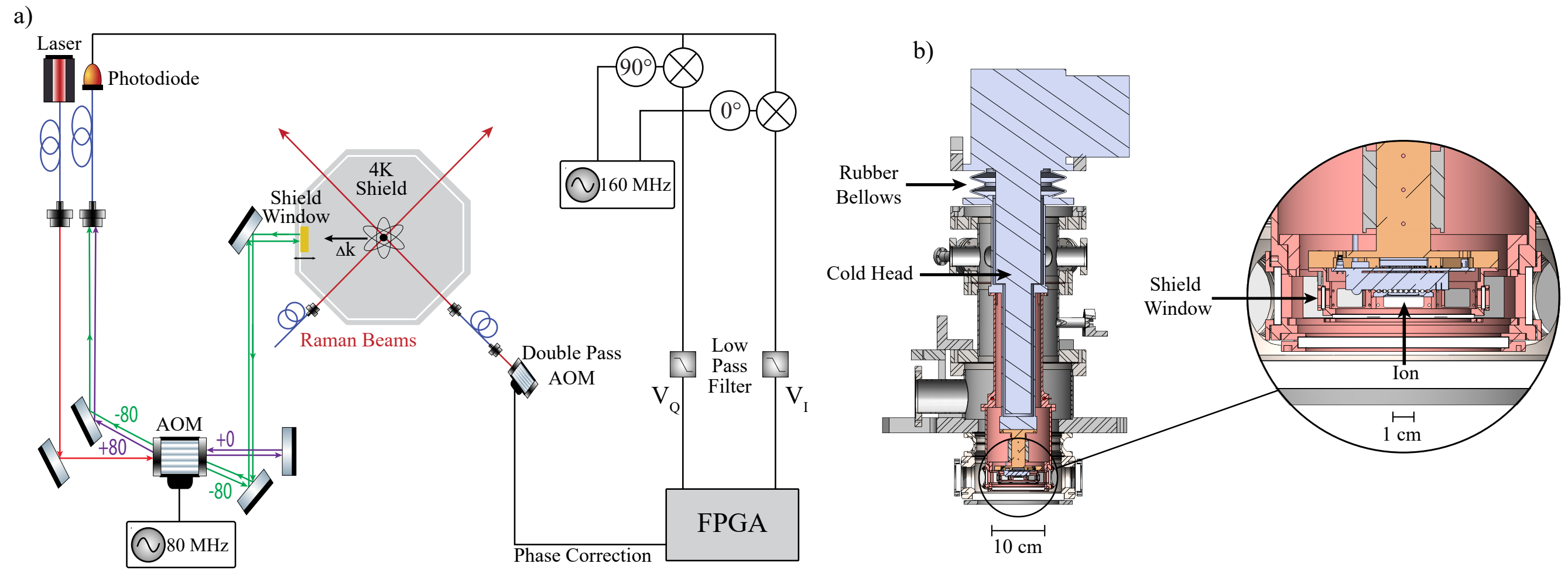}
    \vspace{-7mm}
    \caption{a) Schematic of the vibration-cancellation apparatus where \textbf{$\Delta$k} is the effective wavevector of the Raman beams. The interferometer laser beam is split into a signal arm (green) and a reference arm (purple) that receive different frequency shifts from an AOM driven at 80~MHz. The two arms interfere on the photodiode, and the resulting \qty{160}{\mega\hertz} beat is demodulated with a local oscillator at 160~MHz phase-locked to the AOM driver. An FPGA unwraps the phase, computes the required correction, and updates the RF drive of the double-pass Raman AOM. b) Cross-sectional view of the vibrating cold head with an enlarged view of the vacuum chamber and ion trap.}
    \label{fig:schematic}
\end{figure*}
\subsection{Cryocooler}
 The ion trapping apparatus has been described previously, \cite{sage2012, chiaverini2014, debry2026} but selected components that are especially relevant to this work are discussed in greater detail here and illustrated in Fig.~\ref{fig:schematic}b. Our apparatus uses a closed-cycle cryocooler with two stages of radiation shielding, held at approximately \qty{50}{\kelvin} and \qty{4}{\kelvin}, to thermally isolate the surface-electrode linear Paul trap while maintaining optical access for laser cooling, state preparation, and readout.  The cryogenic system consists of a GM cryocooler, which is mechanically decoupled from the vacuum chamber through flexible rubber bellows and a helium exchange gas space (EGS).  A copper cold finger extends into the vacuum chamber and is thermally coupled to the inner radiation shield via the EGS. The ion trap is installed in a carrier assembly that is attached to the inner radiation shield. 
 
 The cryocooler operates by periodically actuating a displacer piston, generating a quasi-periodic mechanical disturbance that propagates into the vacuum chamber and shield assembly, despite the significant decoupling afforded by the rubber bellows. Hence, the inner shield experiences broad-spectrum vibrations from the cryocooler, both at the GM cycle frequency near 1.2 Hz and at the chamber resonances it excites. Because the trap is rigidly connected to the shield, the ion undergoes the same displacement. We monitor the approximate ion position using one of the inner shield windows, partially coated with gold, as a reference point. No detectable light from the interferometer reaches the ion. 

\subsection{Displacement Interferometer}

The vibrations are measured and corrected using a separate laser system that is not used for qubit operations. We accomplish this with a two-arm heterodyne interferometer that monitors the displacement of the partially metalized window on the inner shield, combined with an FPGA-based feed-forward system that applies a compensating phase correction to the Raman beams used to drive gate operations. A schematic of the setup is shown in Fig.~\ref{fig:schematic}a. The interferometer laser, which operates at a convenient known wavelength, transmits through a dedicated AOM that splits the beam into two spatially and spectrally separated paths. In this work, we use two wavelengths available in our laboratory, 729~nm and 895~nm, highlighting the broadly applicable nature of this technique. The signal arm shown in green in Fig.~\ref{fig:schematic}a receives a \qty{-80}{\mega\hertz} frequency shift on its first pass through the AOM. The beam then reflects off the partially metalized window along the same axis as the effective wavevector $\Delta\mathbf{k} = \mathbf{k}_1 - \mathbf{k}_2$ defined by the Raman laser beams. The round-trip path length changes with the vibrations. On the return pass through the AOM, the signal arm acquires no additional frequency shift, as the diffracted beam is not used. The reference arm shown in purple in Fig.~\ref{fig:schematic}a receives no frequency shift on its first pass through the AOM. It is reflected off a stationary mirror, and on its second pass through the AOM, it receives a +\qty{80}{\mega\hertz} frequency shift. The two arms interfere on a photodiode, creating a heterodyne beatnote at \qty{160}{\mega\hertz} with the differential optical path length encoded in the instantaneous phase.

\subsection{Signal Processing}

The photodiode beatnote is demodulated in quadrature and phase-unwrapped on an FPGA to recover a continuous ambiguity-free record of the window displacement. The signal from the photodiode is split into two paths and mixed with a reference local oscillator (LO) signal with a relative phase shift of $90^{\circ}$ to produce an in-phase $(V_I)$ and quadrature $(V_Q)$ signal. A low-pass filter ($f_{\mathrm{c}}=$~\qty{1.9}{\mega\hertz}) on each path suppresses higher-frequency components. Using an FPGA, we extract an instantaneous phase $\phi$ from the $V_I$ and $V_Q$ components of the heterodyne beatnote using the two-argument arctangent function $\phi = \text{atan}2(V_Q,V_I)$.

The window displacement $d$ along the interferometer axis, for normal incidence, is related to the measured phase by
\begin{equation}
  d = \frac{\phi\,\lambda_\mathrm{int}}{4\pi},
\end{equation}
where $\lambda_\mathrm{int}$ is the interferometer wavelength. Qubit operations are driven by two Raman beams intersecting at approximately $90^\circ$, whose effective wavevector $\Delta\mathbf{k}$ is aligned along the linear trap axis. The beams have a wavelength $\lambda_\mathrm{phys}={859}$~nm and are detuned by approximately \qty{13}{\mega\hertz} from each other and $\Delta\approx$\qty{-1.9}{\tera\hertz} from the $D_{5/2}-P_{3/2}$ transition, giving $|\Delta\mathbf{k}|=\sqrt{2}\,(2\pi/\lambda_\mathrm{phys})$ and $\lambda_\mathrm{Raman}=\lambda_\mathrm{phys}/\sqrt{2}\approx 607$~nm.
A displacement $d$ along $\Delta\mathbf{k}$ shifts the relative phase of the Raman beams at the ion by $2\pi d/\lambda_\mathrm{Raman}$. The required correction phase is therefore
\begin{equation}
  \phi_\mathrm{corr} = \frac{2\pi d}{\lambda_\mathrm{Raman}}
  = \frac{\sqrt{2}\,\lambda_\mathrm{int}}{2\,\lambda_\mathrm{phys}}\,\phi .
  \label{eq:phicorr}
\end{equation}

The correction is applied with the double-pass AOM that modulates one of the Raman beams, so the diffracted light acquires the phase shift twice. The FPGA therefore writes an RF phase
\begin{equation}
  \phi_\mathrm{RF} = \tfrac{1}{2}\phi_\mathrm{corr}
  = \frac{\sqrt{2}\,\lambda_\mathrm{int}}{4\,\lambda_\mathrm{phys}}\,\phi
\end{equation}
 to the DDS channel driving this AOM, corresponding to $\phi_\mathrm{RF}=0.300\,\phi$ for $\lambda_\mathrm{int}=729$~nm and $0.368\,\phi$ for $\lambda_\mathrm{int}=895$~nm. The correction is applied to the Raman beam represented by $\mathbf{k}_1$ in $\Delta\mathbf{k}$. The sign is chosen so that $\phi_\mathrm{corr}$ cancels the phase $\Delta\mathbf{k}\cdot\mathbf{r}(t)$ the ion acquires relative to the two Raman beams, where $\mathbf{r}(t)$ is the ion position.

\subsection{Displacement and Noise Characterization}
\begin{figure}[h]
     \centering
     \includegraphics[width=1\linewidth]{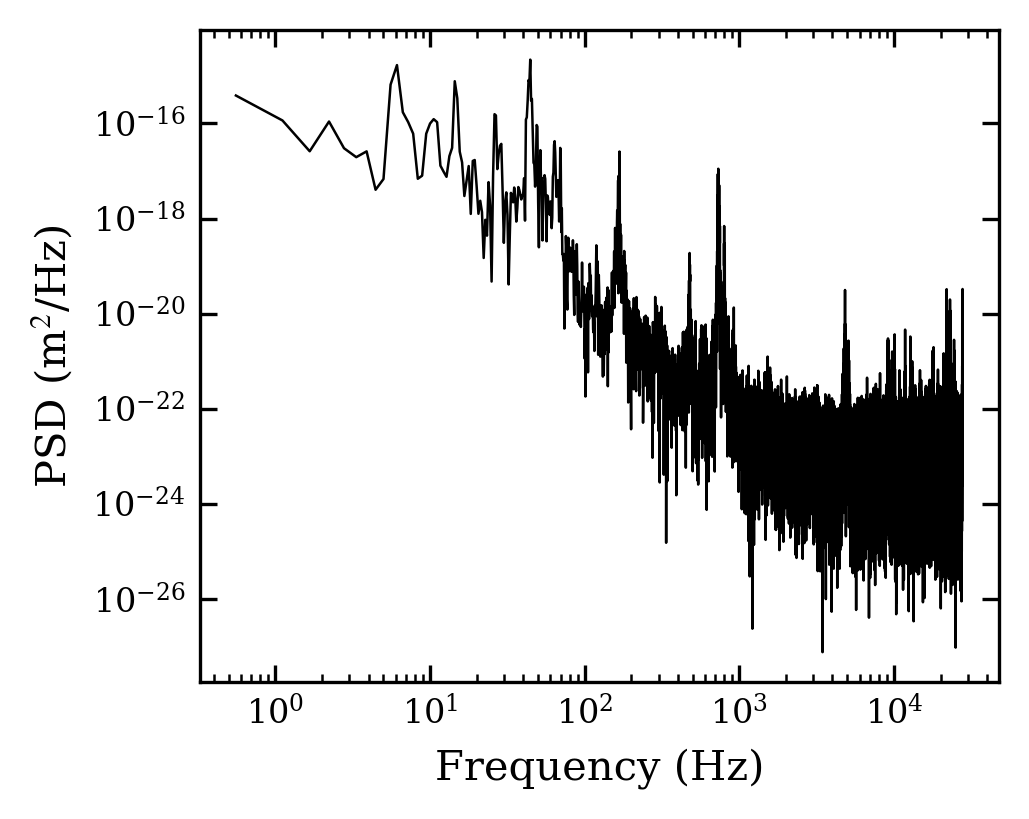}
     \vspace{-7mm}
     \caption{Power spectral density of the inner shield window displacement measured with the interferometer. All significant contributions occur at frequencies below \qty{1}{\kilo\hertz}. This data was taken with an interferometer
     laser at 895 nm.}
     \label{fig:displacement}
\end{figure}

The displacement originates from the GM cryocooler; its quasi-periodic, approximately 1.2~Hz cycle drives the apparatus, and the chamber and shield assembly respond at their mechanical resonances, so that the ion experiences a broad spectrum of vibration frequencies. Fig.~\ref{fig:displacement} shows the power spectral density (PSD) of the interferometer-measured displacement, computed from a \qty{1.8}{\second} record sampled at 55.6~kHz. The RMS displacement over this record is \qty{0.08}{\micro\meter}. The displacement signal was mean-subtracted and weighted by a Hann (cosine squared) window over the full time-series duration to reduce spectral leakage from the finite measurement time. It was then transformed using the real-valued fast Fourier transform (FFT) to obtain the one-sided PSD, normalized by the sampling frequency and window power, giving a frequency resolution of 0.56~Hz.

\section{Experimental Design}\label{experiment}

\subsection{Ion Encoding}
A single $^{40}$Ca$^+$ ion is loaded from a 2D magneto-optical trap of cold calcium atoms and confined \qty{50}{\micro\meter} from the surface of a cryogenic surface-electrode Paul trap, with an axial frequency of \qty{1.9}{\mega\hertz}. The trap is held near \qty{6}{\kelvin}, which suppresses motional heating and yields long trap lifetimes compared to those observed in room-temperature systems.\cite{colin} Prior to each experimental sequence, the axial mode is cooled via Doppler and resolved-sideband cooling to near the motional ground state. State readout is implemented via state-dependent fluorescence detection using 397~nm laser light resonant with the $S_{1/2}$--$P_{1/2}$ transition. Coherence-time measurements are performed using Ramsey spectroscopy of two Zeeman sublevels in the metastable $D_{5/2}$ manifold of $^{40}$Ca$^+$. The two states used in the Ramsey sequence are chosen carefully to ensure that the various $\pi/2$ pulses used in state preparation and analysis do not drive population into unwanted Zeeman sublevels. To quantify the efficacy of the vibration-cancellation scheme, we create the superposition state  
\begin{equation}
    \ket{\psi} = \frac{1}{\sqrt{2}}\bigg (\Ket{m = -\tfrac{5}{2},\, n =1} + \ket{m = -\tfrac{1}{2},\, n = 0}\bigg)
\label{eq:superposition}
\end{equation}
where \textit{m} is the magnetic quantum number and \textit{n} is the axial Fock number.

\begin{figure*}
    \centering
    \includegraphics[width=1\linewidth]{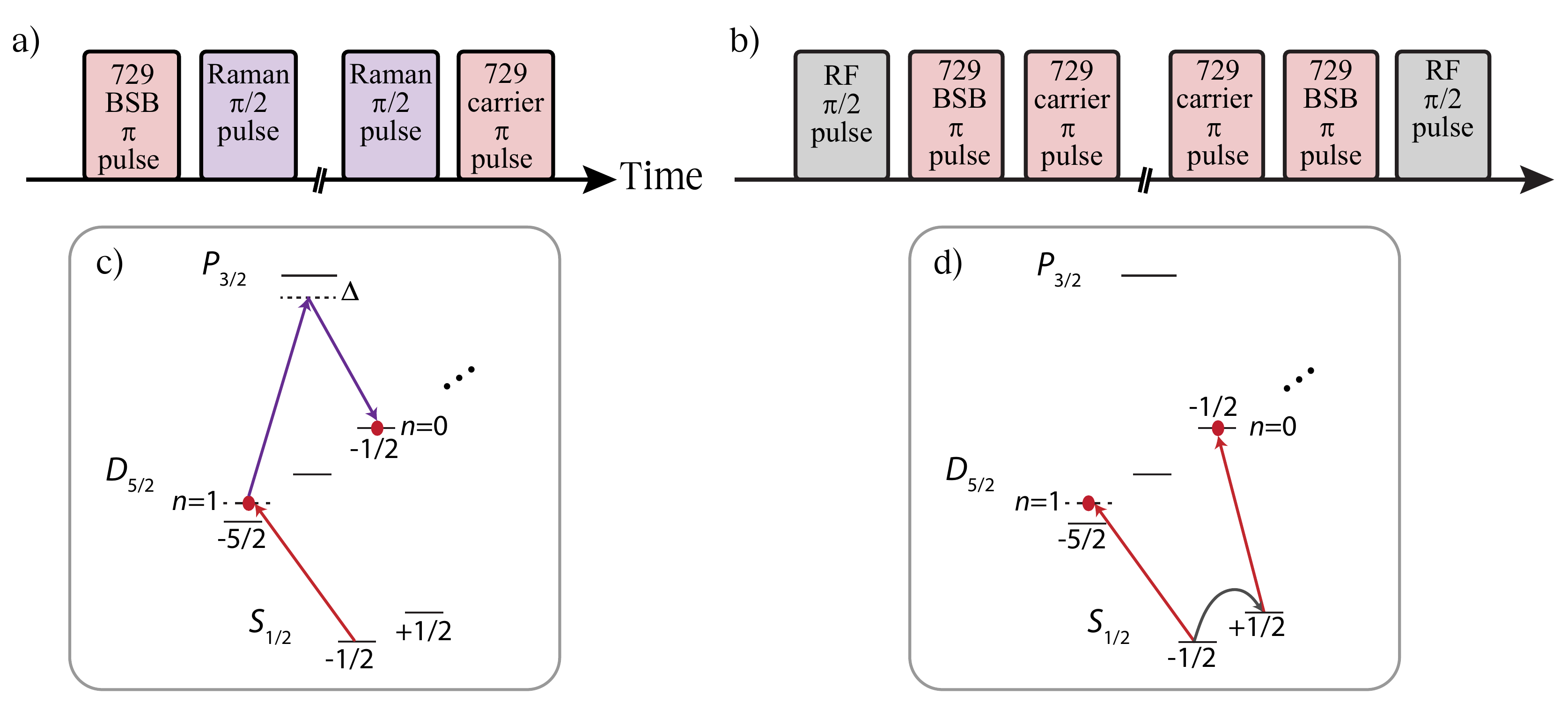}
    \caption{a) Ramsey pulse sequence for the vibration-sensitive coherence-time measurement used to characterize the effect of vibration cancellation. b) Ramsey pulse sequence for the vibration-insensitive coherence-time measurement. c) Partial level structure of $^{40}\mathrm{Ca}^{+}$ showing the transitions that prepare the Ramsey superposition for the vibration-sensitive Ramsey sequence, with the color of the arrows matching the corresponding pulse in a). Here $n$ is the axial Fock number and $\Delta$ is the detuning from the ${P} _{3/2}$ state. The two purple arrows represent a single Raman $\pi/2$ pulse. d) Partial level structure of $^{40}\mathrm{Ca}^{+}$ showing the transitions used to prepare the vibration-insensitive Ramsey superposition, with the colors of the arrows corresponding to the colors of the pulses in b).}
    \label{fig:energylevel}
\end{figure*}

\subsection{Pulse Sequences}

The two pulse sequences shown in Fig.~\ref{fig:energylevel} share the same Ramsey structure and readout procedure but generate the superposition state through different transitions: one sensitive to the Raman beams' effective $k$-vector and therefore vibration-induced phase noise, and one insensitive to it, therefore providing a vibration-free baseline to judge the performance of the cancellation system. 

The pulse sequence used to measure the vibration-sensitive coherence
time is shown in Fig.~\ref{fig:energylevel}a. Doppler cooling and resolved-sideband
cooling prepare the ion in the
$\Ket{S_{1/2},m{=}{-}\tfrac{1}{2}, n{=}0}$ state. A
blue-sideband (BSB) $\pi$ pulse at 729~nm transfers the population to $\Ket{D_{5/2},m{=}{-}\tfrac{5}{2}, n{=}1}$. The
first Raman $\pi/2$ pulse (duration $\sim$\qty{21}{\micro\second}) creates the superposition of Eq.~(\ref{eq:superposition}) by using an intermediate virtual $P_{3/2}$ state as shown in Fig.~\ref{fig:energylevel}c. During the free-evolution (Ramsey delay) time~$T$, the superposition accumulates a relative phase. With vibration cancellation on, the interferometer measures the window displacement and applies a single phase correction to the second Raman $\pi/2$ pulse based on net displacement. Finally, a 729~nm carrier laser $\pi$ pulse returns the $\Ket{ m{=}{-}\tfrac{5}{2}}$ population to the $S_{1/2}$ state, after which the ion's fluorescence is measured with the 397~nm laser applied to distinguish the two states.

The pulse sequence for the vibration-insensitive coherence-time
measurement is shown in Fig.~\ref{fig:energylevel}b. Doppler cooling and resolved sideband cooling prepare $\Ket{S_{1/2},m{=}{-}\tfrac{1}{2}, n{=}0}$. A $\pi/2$ RF pulse on the Zeeman ($S_{1/2}$) qubit transition at approximately 12~MHz creates the initial equal superposition of the
$\Ket{ m{=}{-}\tfrac{1}{2}, n{=}0}$ and $ \Ket {m{=}{+}\tfrac{1}{2}, n{=}0}$ states. A 729~nm laser pulse sequence consisting of a BSB $\pi$ pulse and a carrier $\pi$ pulse maps the $S_{1/2}$ Zeeman sublevel populations into the $\Ket{ m{=}{-}\tfrac{5}{2}, n{=}1}$ and $\Ket{ m{=}{-}\tfrac{1}{2}, n{=}0}$ states of the $D_{5/2}$ manifold. During the free-evolution time, the superposition accumulates a phase that, unlike the pulse sequence in Fig.~\ref{fig:energylevel}a, is insensitive to vibrations. The BSB and carrier laser pulses are then repeated in reverse to map the state back onto the $S_{1/2}$ manifold. A second $\pi/2$ RF pulse closes the Ramsey sequence. Finally, a 729~nm shelving pulse transfers the population of a single $S_{1/2}$ sublevel to the $D_{5/2}$ state to allow for state-dependent fluorescence detection. 

Measurement and phase-update latencies limit the vibration-noise frequencies this scheme can address. We determine the correctable frequency range by characterizing the dominant delays in the control loop. There are \qty{23}{\micro\second} of dead time from the center of the first Raman $\pi$/2 pulse (Fig.~\ref{fig:energylevel}a) to the end of the first phase measurement during the Ramsey delay. At the end of the Ramsey delay, there are approximately \qty{29}{\micro\second} of dead time between the start of the last phase measurement and the center of the second Ramsey $\pi$/2 pulse. The additional dead time results from updating the phase of the second pulse. Vibrations that evolve during these intervals are not measured and therefore are not corrected. For a sinusoidal vibration at frequency $f$, a dead time $t_d$ leaves an uncorrected phase of approximately $2\pi f t_d$ times the vibration-induced phase amplitude. The errors from the two intervals have a relative phase that varies with the Ramsey delay and therefore add approximately in quadrature, giving an effective dead time $t_\mathrm{eff}\approx$~\qty{37}{\micro\second}. The correction is thus effective for $f\ll1/(2\pi t_\mathrm{eff})\approx$~\qty{4.3}{\kilo\hertz}, and the uncorrected fraction is approximately 0.023 at 100~Hz and 0.23 at 1~kHz.


\section{Results}\label{results}

\subsection{Coherence Time With and Without Vibration Cancellation}
\begin{figure}[h]
    \centering
    \includegraphics[width=1\linewidth]{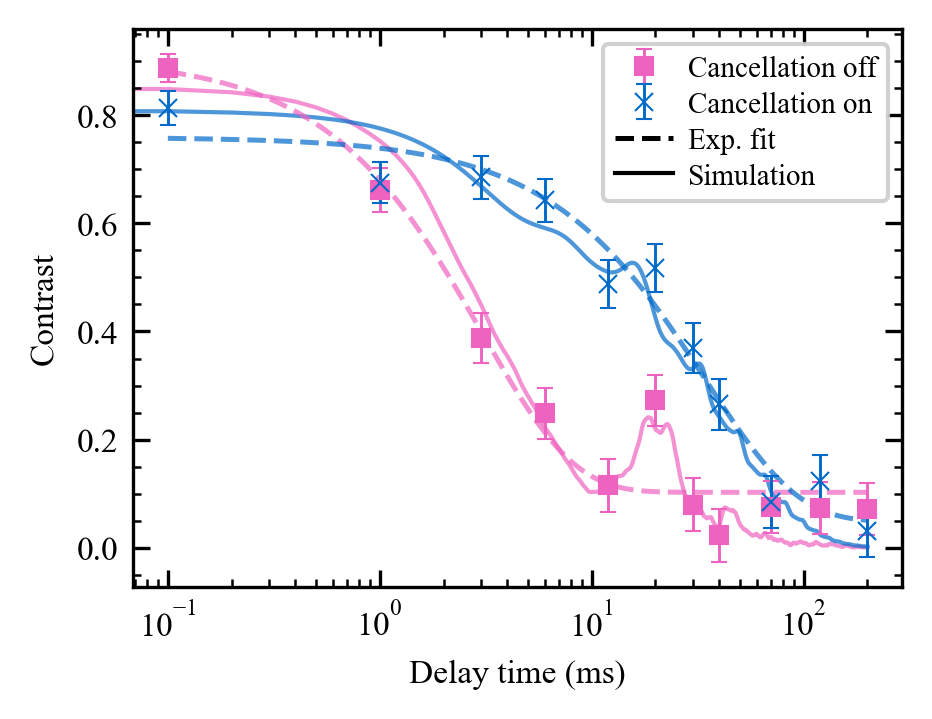}
    \vspace{-7mm}
    \caption{Contrast as a function of delay time of the metastable-qubit Ramsey experiment with (blue) and without (pink) vibration cancellation. Data points with error bars, exponential fits (dashed lines), and coherence time simulations (solid line) with (pink) and without (blue) the measured vibration noise are shown. This data was taken with an interferometer laser at \qty{729}{\nano\meter}. Error bars are one-standard-deviation statistical uncertainties in the
contrast, obtained from the sinusoidal fit to each Ramsey phase scan.}
    \label{fig:vib no vib}
\end{figure}

 Figure~\ref{fig:vib no vib} shows the measured Ramsey fringe contrast as a function of free-evolution time $T$, spanning \qty{0.1}{\milli\second} to \qty{200}{ms}, with and without the vibration-cancellation system active. The Ramsey contrast is extracted by scanning the phase of the second Ramsey pulse and fitting a sinusoid to the resulting fringe. Vibration cancellation results in an order-of-magnitude improvement in the qubit coherence time. The coherence times are extracted from an exponential fit to the Ramsey contrast decay of the form
\begin{equation}
    C(T) = A \exp\!\left[-\frac{T}{\tau}\right] + C_0,
    \label{eq:exp}
\end{equation}
where $A$ is the fitted contrast amplitude, $\tau$ is the $1/e$ coherence time, and $C_0$ is a constant offset accounting for residual contrast at long delay. A fully dephased Ramsey signal has zero contrast; the positive offset $C_0$ instead reflects a statistical bias of the contrast estimator: here $C_{0,\text{off}}=0.10\pm0.02$ and $C_{0,\text{on}}=0.049\pm0.037$. For cancellation off, the fit yields $\tau_{\text{off}} = 3.0 \pm 0.4$~ms, while for cancellation on, the fit yields $\tau_{\text{on}} = 35 \pm 5$~ms. The solid lines in Fig.~\ref{fig:vib no vib} show a coherence time simulation with (pink) and without (blue) the measured vibration noise.

To better understand noise processes in our system, we perform simulations of the effect of the vibrations and other decoherence sources. The simulations incorporate three independent measurements of noise sources: the measured vibration data, measurements of the Raman beams' relative phase noise, and measurements of the motional coherence time of the system. The vibration noise was modeled using the data shown in Fig. \ref{fig:displacement}. The Raman beams' relative phase fluctuations that occur upstream of the chamber, which arise due to changes in the effective path lengths of optical fibers and are not cancelled by the system described here, were independently measured by comparing the beatnote of an interferometer constructed after the optical fibers with a quartz oscillator reference. Finally, the decoherence of the different motional Fock states used in the metastable-state Ramsey measurements was measured independently with the pulse sequence shown in Fig. \ref{fig:energylevel}b and modeled by adding a white noise term with a \qty{40}{ms} dephasing time. 

To find the decoherence rate under different conditions, we numerically calculate the dynamics of a two-level system initially prepared in an equal superposition of the two states and simulate the effect of the vibrations as a Doppler shift that applies an energy shift to the higher-lying state. We simulate averaged behavior over many experimental trials by performing Monte-Carlo samples using measured vibration trajectories. The output of the simulation is shown in Fig.~\ref{fig:vib no vib} with all noise sources included (pink) and with the vibrations excluded (blue). The simulation curves shown include only one free parameter: a maximum contrast (due to state preparation errors not captured by the simulations).

The agreement of these simulations with our experimental data suggests that we are able to eliminate vibrations as a dominant source of decoherence in the experiment. When we include all decoherence sources, the simulations agree well with the experimental measurements with the vibration-cancellation system disabled, including a slight contrast recovery around \qty{20}{\milli\second} corresponding to a Fourier spectrum peak near \qty{50}{\hertz}. When we exclude the measured vibration noise from the simulation, it matches well with the experimental data with the vibration-cancellation system active, suggesting that any residual effects of vibrations on the qubit coherence are minimal.

\subsection{Decoherence Rates as a Function of Vibration Strength}

\begin{figure}[h]
    \centering
    \includegraphics[width=1\linewidth]{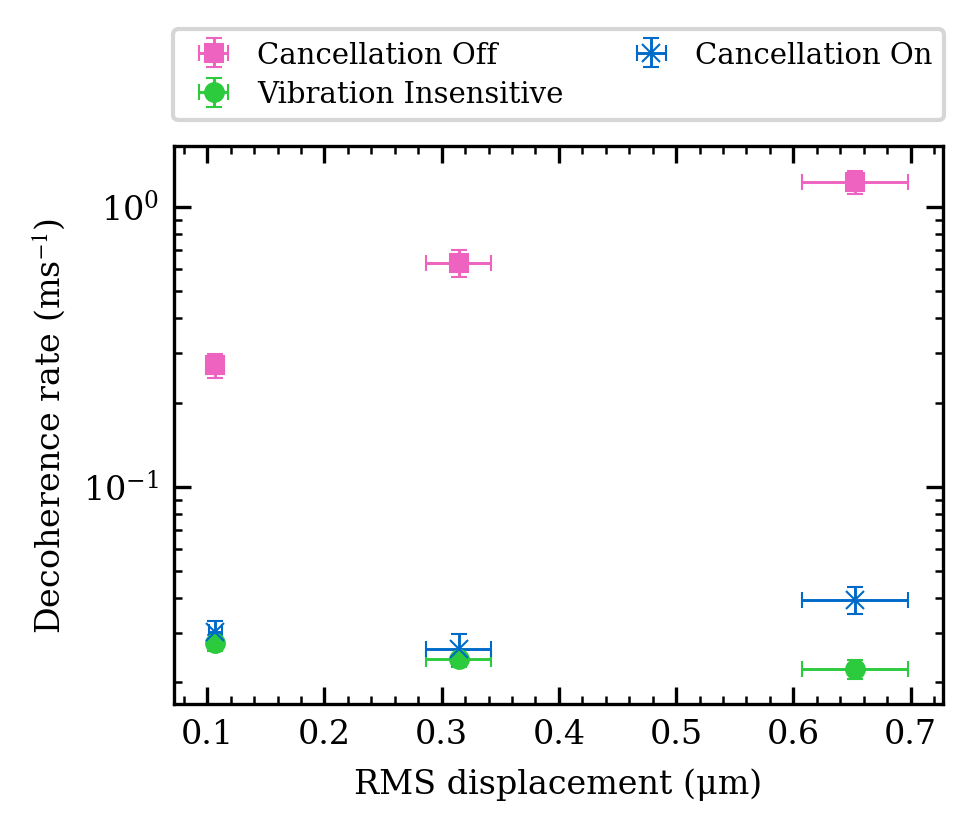}
    \vspace{-7mm}
    \caption{Decoherence rate (ms$^{-1}$) as a function of vibration strength, characterized by the RMS displacement~(\qty{}{\micro\meter}) for three cases: cancellation off, cancellation on, and the same state prepared in a vibration-insensitive way (no Raman beams). This data was taken with an interferometer laser at \qty{895}{\nano\meter}.  Vertical error bars on the decoherence rate are one-standard-deviation uncertainties
    propagated from the fitted 1/e coherence time $\tau$ of Eq.~(5). Horizontal error bars are the standard error of the RMS displacement, estimated by block averaging the interferometer record. Some error bars are smaller than the markers.}
    \label{fig:vib coupling}
\end{figure}

To further characterize the vibration-cancellation system, we perform a set of experiments that artificially increase the amplitude of the vibrations coupled to the chamber\cite{niffenegger2020} by bypassing the bellows shown in Fig.~\ref{fig:schematic} with 3D-printed rectangular polylactic acid (PLA)  plates that connect the vibrating cryocooler to the chamber. Larger vibration amplitudes are achieved by increasing the number of coupling plates. Figure~\ref{fig:vib coupling} shows the decoherence rates observed as a function of applied vibration amplitude, characterized by the RMS displacement, for three cases: cancellation off, cancellation on, and the vibration-insensitive reference. The decoherence rate with active cancellation stays near the reference for the native coupling (RMS displacement: 0.106~$\pm$~\qty{0.005}{\micro\meter}) and one plate of coupling (RMS displacement: 0.31~$\pm$~\qty{ 0.02}{\micro\meter}) and degrades at two plates of coupling (RMS displacement: 0.65~$\pm$~\qty{ 0.04}{\micro\meter}), while the uncancelled decoherence rate rises monotonically. As expected, the vibration-insensitive measurement is not affected by the increased magnitude of vibrations.

Because the residual decoherence rate with cancellation active is set by nonvibrational sources, the improvement in coherence time does not directly measure cancellation performance. We therefore define the vibration-induced rate $\gamma_\mathrm{vib,on/off}=\gamma_\mathrm{on/off}-\gamma_\mathrm{ref}$, where $\gamma_\mathrm{on/off}=1/\tau_\mathrm{on/off}$, $\tau_\mathrm{on/off}$ is the $1/e$ coherence time from the exponential fit, and
$\gamma_\mathrm{ref}$ is the rate of the vibration-insensitive sequence measured at the same coupling. The suppression factor $S$ is given by
\begin{equation}
  S=\frac{\gamma_\mathrm{vib,off}}{\gamma_\mathrm{vib,on}}.
  \label{eq:suppression}
\end{equation}
At RMS displacements of 0.11 and 0.31~\textmu m, $\gamma_\mathrm{vib,on}$ is consistent with zero ($0.7\sigma$ and $0.5\sigma$), so we report lower bounds on $S$ obtained from the $2\sigma$ upper limit on $\gamma_\mathrm{vib,on}$ and the central value of $\gamma_{\rm vib,off}$: $S>26$
and $S>62$, respectively. At 0.65~\textmu m,
$\gamma_\mathrm{vib,off}=1.20\pm0.11$~ms$^{-1}$ and
$\gamma_\mathrm{vib,on}=0.0172\pm0.0047$~ms$^{-1}$, giving $S=70^{+27}_{-16}$.


\section{Discussion and Outlook}\label{discussion}

We have demonstrated an adaptable vibration-cancellation system for trapped-ion qubits that achieves approximately an order-of-magnitude improvement to coherence time and an approximately 70-fold suppression of vibration-induced decoherence at an RMS displacement of 0.65~\textmu m. The system measures the displacement of the inner shield window with a two-arm interferometer and applies a compensating phase correction to one of the Raman beams, suppressing the dominant decoherence mechanism imposed by the closed-cycle cryocooler. 

With active cancellation, vibrations are no longer the limiting factor for the metastable-qubit coherence investigated here. The residual coherence time after cancellation is consistent with that of the vibration-insensitive reference state at native coupling (Fig.~\ref{fig:vib coupling}), demonstrating that the cancellation system suppresses vibration-induced decoherence below other decoherence mechanisms. This suppression persists even when vibration amplitudes are increased above those produced by the cryocooler alone (Fig.~\ref{fig:vib coupling}), suggesting robustness to increased vibration amplitude. Additionally, the level of agreement between the measured and simulated coherence times (Fig.~\ref{fig:vib no vib}) indicates that the cancelled case reaches the nonvibrational floor. This is consistent with the displacement spectrum in Fig.~\ref{fig:displacement}; all significant vibration components lie below \qty{1}{\kilo\hertz}, well under the \qty{4.3}{\kilo\hertz} correction limit set by the dead time, so the uncorrected fraction is at most 0.23 and far smaller for the dominant low-frequency components. Residual limitations are likely dominated by motional decoherence and magnetic-field decoherence, which are independent of cryocooler vibrations. 

The primary sources of residual error in the cancellation are: (i) photon shot noise on the interferometer photodiode, (ii) the geometric approximation that the vibration of the inner shield window equals the ion displacement along $\Delta\mathbf{k}$, (iii) misalignment of the interferometer axis from $\Delta\mathbf{k}$, and (iv) the motion of the interferometer reference relative to the Raman delivery optics, which is written onto the qubit. The good agreement between the measured post-cancellation coherence times and the vibration-insensitive reference suggests these residual errors are small compared with other decoherence mechanisms in our system. However, for amplitudes larger than the native vibration amplitudes, dead time is likely to become a more significant error source. 

The design is broadly applicable to trapped-ion systems operating with a cryocooler or any other source of periodic mechanical noise. Although it requires optical access along the $\Delta\mathbf{k}$ axis, it can be retrofitted onto any existing trapped-ion apparatus with a metalized inner shield viewport and basic optics. We use commercially available ion-control hardware (a Sinara crate running ARTIQ (M-Labs) with a Sinara 1125 Processor “Kasli-SoC”, Sinara 5108 Sampler, and Sinara 4410 DDS “Urukul” AD9910 cards) to calculate the compensating phase signal, but a low-cost independent FPGA could be used instead \cite{shaw2019} and would reduce dead time. Additional $\Delta\mathbf{k}$ directions could be addressed by adding interferometers along other axes. The FPGA-based signal-processing pipeline is adjustable to different interferometer light frequencies, as we have demonstrated by the use of both \qty{729}{\nano\meter} and \qty{895}{\nano\meter} lasers in Figs.~\ref{fig:vib no vib} and \ref{fig:vib coupling}. This approach is a general-purpose tool for extending coherence times in cryogenic trapped-ion systems that use GM or pulsed cryocoolers.


\section*{Acknowledgment}
DISTRIBUTION STATEMENT A. Approved for public release. Distribution is unlimited. This material is based upon work supported by the U.S. Army Research Office through grant W911NF-24-1-0379 and by the Under Secretary of War for Research and Engineering under Air Force Contract No. FA8702-15-D-0001 or FA8702-25-D-B002. Support is also acknowledged from the U.S. Department of Energy, Office of Science, National Quantum Information Science Research Centers, Quantum Systems Accelerator (Award No. DE-SCL0000121). I.L.C. acknowledges support in part by the NSF Center for Ultracold Atoms. Any opinions, findings, conclusions or recommendations expressed in this material are those of the author(s) and do not necessarily reflect the views of the Department of Energy and Under Secretary of War for Research and Engineering.

\section*{Author Declarations}
Alexandra Plocki: Conceptualization (equal); Investigation (lead); Formal analysis (lead); Writing – original draft (lead). Kyle DeBry: Conceptualization (equal); Investigation (supporting), Writing – original draft  (supporting). Agustin Valdes Martinez: Investigation (supporting).  Isaac L. Chuang: Supervision (equal). Robert McConnell: Supervision (equal). Colin D. Bruzewicz: Supervision (equal). The authors have no conflicts to disclose.
\section*{Data Availability}
The data that support the findings of this study are available from the corresponding author upon reasonable request.
\section*{References}

\bibliography{bib}

\end{document}